\documentclass[aps,twocolumn,10pt,letterpaper,superscriptaddress]{revtex4-1}
\usepackage{amsmath}
\usepackage{graphicx}
\newcommand{\bra}[1]{\left< #1 \right\vert}
\newcommand{\ket}[1]{\left\vert #1 \right>}

\begin{document}
\title{Multiphoton Discrete Fractional Fourier Dynamics in Waveguide Beam Splitters}

\author{Konrad Tschernig}\email{konrad.tschernig@physik.hu-berlin.de}
\affiliation{Humboldt-Universit\"at zu Berlin, Institut f\"ur Physik, AG Theoretische Optik $\&$ Photonik, 12489 Berlin, Germany}
\author{Roberto de J. Le\'on-Montiel}
\affiliation{Instituto de Ciencias Nucleares, Universidad Nacional Aut\'onoma de M\'exico, 70-543, 04510 Cd. Mx., M\'exico}
\author{Omar S. Maga\~na-Loaiza}
\affiliation{National Institute of Standards and Technology, 325 Broadway, Boulder, Colorado 80305, USA}
\author{Alexander~Szameit}
\affiliation{Institut f\"ur Physik, Universit\"at Rostock D18051 Rostock, Germany}
\author{Kurt Busch}
\affiliation{Humboldt-Universit\"at zu Berlin, Institut f\"ur Physik, AG Theoretische Optik $\&$ Photonik, 12489 Berlin, Germany}
\affiliation{Max-Born-Institut, Max-Born-Stra\ss{}e 2A, 12489 Berlin, Germany}
\author{Armando Perez-Leija}
\affiliation{Humboldt-Universit\"at zu Berlin, Institut f\"ur Physik, AG Theoretische Optik $\&$ Photonik, 12489 Berlin, Germany}
\affiliation{Max-Born-Institut, Max-Born-Stra\ss{}e 2A, 12489 Berlin, Germany}
\date{\today}

\begin{abstract}
We demonstrate that when a waveguide beam splitter (BS) is excited by N indistinguishable photons, the arising multiphoton states evolve in a way as if they were coupled to each other with coupling strengths that are identical to the ones exhibited by a discrete fractional Fourier system. Based on the properties of the discrete fractional Fourier transform, we then derive a multiphoton suppression law for 50/50 BSs, thereby generalizing the Hong-Ou-Mandel effect. Furthermore, we examine the possibility of performing simultaneous multiphoton quantum random walks by using a single waveguide BS in combination with photon number resolving detectors. We anticipate that the multiphoton lattice-like structures unveiled in this work will be useful to identify new effects and applications of high-dimensional multiphoton states.
\end{abstract}
\maketitle
\cleardoublepage
\section{Introduction}
\begin{figure}[t!]
\begin{center}
\includegraphics[width=0.5\textwidth]{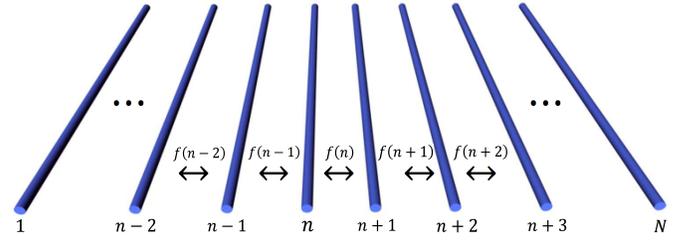}
\caption{(a) Schematic of a waveguide lattice aimed to perform discrete fractional Fourier transforms of optical signals sampled at $N$ data points.
The system consists of $N$ identical evanescently coupled waveguides having coupling coefficients that obey the function $f(n)=\frac{\kappa}{2}\sqrt{n(N-n)}$.}
\label{fig:array}
\end{center}
\end{figure}

\begin{figure}[t!]
\begin{center}
\includegraphics[scale=0.4]{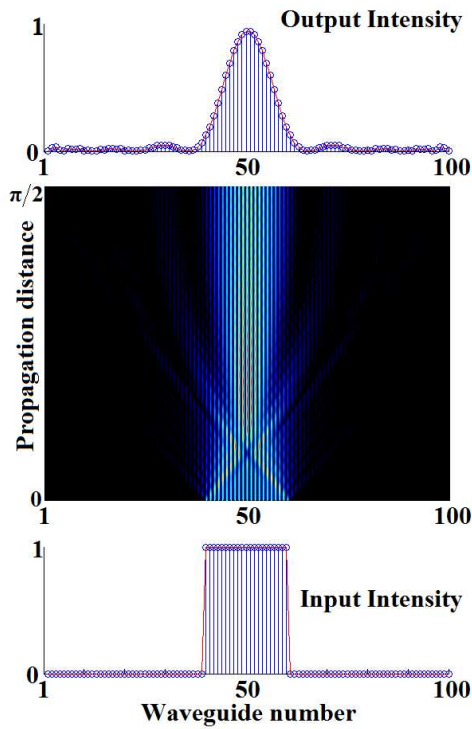}
\caption{Transformation of a discrete rectangular function into a sinc function using a photonic lattice of $100$ waveguides and scaling factor $\kappa=1$. At the bottom we show the initial intensity profile of a rectangular function sampled at $21$ data points. In the central panel we present the intensity evolution in the range $z\in[0,\pi/2]$ which represents all fractional orders of the FT of the initial field. The upper figure depicts the usual FT obtained at the output of the system, i.e., at $z=\pi/2$.}
\label{fig:evolution}
\end{center}
\end{figure}

\setlength{\belowcaptionskip}{-20pt}
In 1980 Namias introduced the continuous fractional Fourier transform (FrFT) as a new mathematical tool for solving quantum mechanical problems \cite{Namias}. The importance of the FrFT lies in the fact that it contains the standard Fourier transform (FT) as a particular case and, as a result, every property of the FT can be generalized to the fractional Fourier domain \cite{Mendlovic,Ozaktas,Lohmann,Ozaktas2}. Over the years, the FrFT has found widespread applicability in different areas such as filter design, signal analysis, phase retrieval protocols, and pattern recognition \cite{Weimin,Lohmann2}.
Concurrently, in the field of optics, where FTs occur naturally, the FrFT has yielded interesting applications ranging from spatial-temporal optical field reconstruction to optical image encryption \cite{Ozaktas2}. \\
Similarly to the standard discrete FT, a discrete version of the FrFT, the so-called discrete FrFT (DFrFT), has been developed to operate over finite discretely sampled signals \cite{Atakishiyev}. Quite recently, such DFrFT has been demonstrated in optics using judiciously engineered waveguide arrays and its applicability for classical and quantum light has been shown \cite{Weimann}. In the classical domain the properties of the DFrFT can be exploited to implement digital medical algorithms \cite{Zhang}, optical field tomography \cite{Manko,McAlister}, and phase estimation protocols \cite{Orlowski}. Correspondingly, in the quantum realm, the DFrFT can be used for the generation of multiple-path NOON states \cite{LeijaJx}, to perform multiple phase estimation protocols \cite{Humphreys}, and potentially to transfer \cite{LeijaJx2} and to store qubits \cite{Weimann}.\\
Notably, in order to implement a fractional Fourier transformer applicable to signals sampled at $N$ data points, we require a system of $N$ coupled sites such as cavities, potential wells, and waveguides, in which the adjacent elements, $n$ and $n+1$, exhibit a coupling strength given by the function $f(n)=\frac{\kappa}{2}\sqrt{n(N-n)}$, with $\kappa$ denoting a scaling factor \cite{Weimann}.\\
For the sake of completeness, we briefly describe the implementation of the DFrFT using waveguide lattices. To do so, we consider the waveguide array shown in Fig.~(\ref{fig:array}). In such photonic systems, a discrete version of the usual FT is produced at the propagation distance $z=\pi/2\kappa$, while any fractional order of the FT is observed at distances within the interval $z\in(0,\pi/2\kappa)$ \cite{Weimann}. To illustrate these facts, let us assume an array of $100$ waveguides excited by an optical field of the same intensity covering the $21$ central waveguides as depicted in Fig.~(\ref{fig:evolution}). Evidently, the input field corresponds to a rectangular function sampled at 21 points, whose FT is a discrete sinc function. In principle, along evolution the optical field undergoes an infinite number of transformations being the FT one particular case, and for any other propagation distance $z\neq\pi/2\kappa$ the discrete fractional Fourier transform arises, see Fig.~(\ref{fig:evolution}).\\
\section{Discrete Fractional Fourier Transform of Multiphoton states}
\begin{figure}[t!]
\begin{center}
\includegraphics[width=0.5\textwidth]{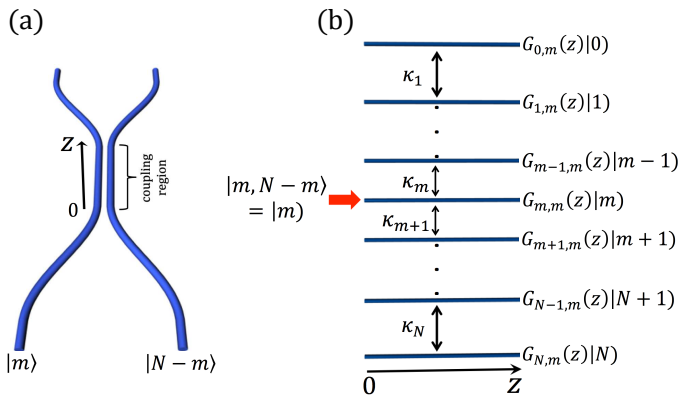}
\caption{(a) Schematic view of a BS consisting of two identical evanescently coupled waveguides having a  coupling coefficient $\kappa/2$. The system is initialized with the $N$-photon state $|m):= \ket{m,N-m}$, representing $m$ photons launched into the left port and $N-m$ photons into the right one. (b) State diagram for the $(N+1)$ multiphoton states propagating in the BS. Every line represents one state which appears to be coupled to the adjacent states, and $G_{n,m}(z)$ is the probability amplitude for the state $|n)=\ket{n,N-n}$.}
\label{fig:coupler}
\end{center}
\end{figure}

\begin{figure}[t!]
\includegraphics[width=0.5\textwidth]{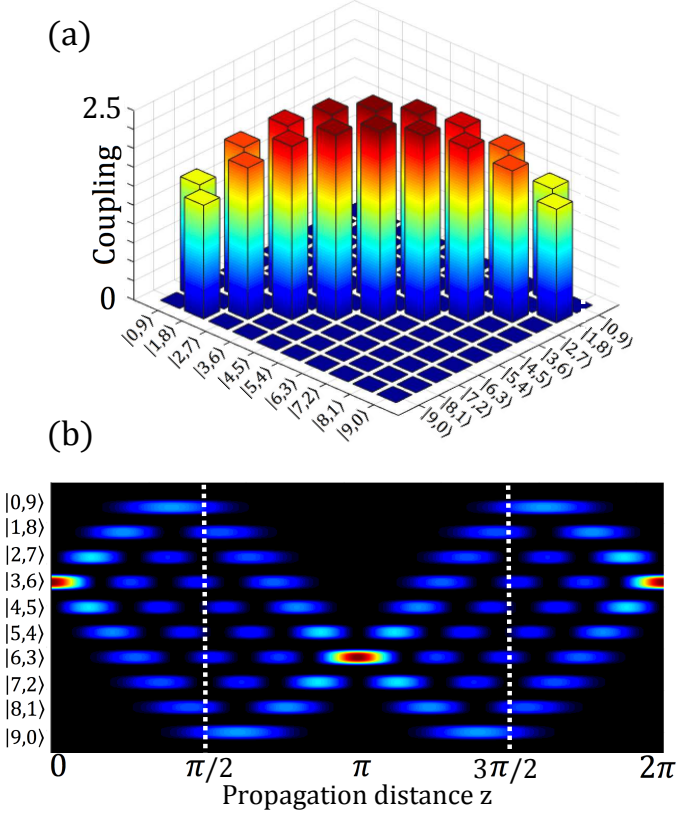}
\caption{(a) Coupling distribution obtained from the matrix elements $\hat{H}_{n,m}$ for $N=9$ photons injected into a waveguide BS with coupling constant $\kappa/2=1/2$. (b) Theoretical probability evolution, $|G_{n,m}(z)|^2$, corresponding to a BS initially excited by the $N=9$ photon state $\ket{\psi(0)}=|3)=\ket{3,6}$.}
\label{fig:Dyn}
\end{figure}

\begin{figure}[b!]
 \includegraphics[scale=0.4]{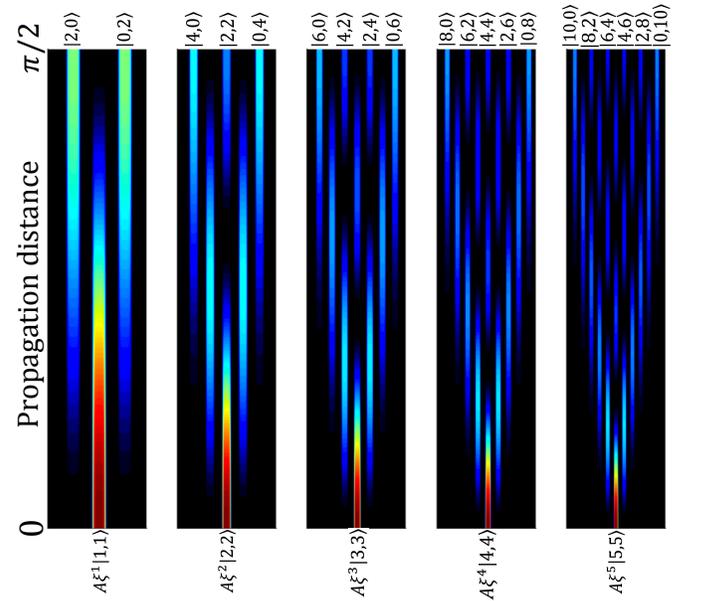}
  \caption{First five non-vanishing probability distributions arising in a waveguide BS initially excited by a TMSVS , $\ket{\xi}=A\sum_{n=0}^{\infty}\xi^{n}\ket{n,n}$, with $A=\sqrt{1-|\xi|^{2}}$, and $\xi=0.5$.}
  \label{fig:parallel}
\end{figure}

In this paper, we theoretically show that $N$ indistinguishable photons injected into a waveguide beam splitter (BS) give rise to multiphoton lattice-like structures where the associated \textit{coupling coefficients} are identical to the ones exhibited by a discrete fractional Fourier system. 
Given the fact that fractional Fourier systems have enabled interesting applications in classical optics, it would be of great interest to study alike processes at the level of multiphoton states. Indeed, the quantum and statistical properties of photons interfering in a BS have brought to light very profound implications such as the Hong-Ou-Mandel interference effect \cite{Hong}, which in turn has opened the door to very intriguing applications in quantum metrology and computing \cite{Walmsley}. As it turns out, the quantum properties of bulk BSs have been throughly explored by many authors \cite{Zeilinger,Ou,Yurke,Campos}, however, none of those investigations has pointed out the occurrence of multiphoton discrete fractional Fourier dynamics in such elementary systems. \\
\noindent
We start our analysis by considering a waveguide BS being excited by $N$ indistinguishable photons prepared in the state $\ket{m,N-m}=\ket{m}\ket{N-m}$, see Fig.~(\ref{fig:coupler}~a). Physically, the state $\ket{m}$ represents $m$ photons being launched into port $a$ (left waveguide in Fig.~(\ref{fig:coupler} a)), while $\ket{N-m}$ stands for $N-m$ photons entering port $b$ (right waveguide in Fig.~(\ref{fig:coupler} a)). Quantum mechanically the BS is governed by the Hamiltonian \cite{Lai}
\begin{equation}\label{eq:Ham}
\hat{H}=\beta \hat{a}^\dagger \hat{a} +\beta\hat{b}^\dagger \hat{b} + \frac{\kappa}{2} \left( \hat{a}\hat{b}^\dagger + \hat{a}^\dagger \hat{b} \right),
\end{equation}
where $\beta$ represents the propagation constant, $\kappa/2$ denotes the coupling coefficient, and $\hat{a}^{\dagger}$, $\hat{b}^{\dagger}$
$(\hat{a},\hat{b})$ are the boson creation (annihilation) operators of modes $a$ and $b$, respectively. 
Typically, the evolution of the single-photon creation operators in waveguide BSs is described by the transformation  \cite{Weedbrook,Dellanno}
\begin{align}
\begin{pmatrix} 
\hat{a}^{\dagger}(z) \\ 
\hat{b}^{\dagger}(z)
\end{pmatrix}=e^{i\beta z}
\begin{pmatrix} \cos\left(\frac{\kappa z}{2}\right) & i\sin\left(\frac{\kappa z}{2}\right) \\ i\sin\left(\frac{\kappa z}{2}\right)  \cos\left(\frac{\kappa z}{2}\right) \end{pmatrix}\begin{pmatrix} 
\hat{a}^{\dagger}(0) \\ 
\hat{b}^{\dagger}(0)
\end{pmatrix}.
\end{align}
Notice, from this transformation one readily infers that to implement a 50/50 BS one has to design the system with a propagation length of $z=\frac{\pi}{2\kappa}$.
To show the connection of the waveguide BS with the DFrFT we take a different approach in which the dynamics induced by Hamiltonian \eqref{eq:Ham} over the $N$-photon states $\ket{m,N-m}$ is computed from the matrix elements $\hat{H}_{n,m}=\bra{n,N-n}\hat{H}\ket{m,N-m}$. To do so, we introduce the compact notation $|m)$ to represent the state $\ket{m,N-m}$ indicating that $m$ photons are injected into port $a$ and $N-m$ photons into port $b$. By applying the Hamiltonian operator Eq.~(\ref{eq:Ham}) to the states $|m)$, one obtains the matrix elements $\hat{H}_{n,m}=N \beta \delta_{n,m}+\kappa_{m} \delta_{n,m-1} + \kappa_{m+1} \delta_{n,m+1}$, where the \textit{effective coupling coefficient} $\kappa_m=\frac{\kappa}{2}\sqrt{m(N+1-m)}$ governs the transition probability between all possible $N$-photon states $\left(|0)=\ket{0,N}, |1)=\ket{1,N-1},..., |N)=\ket{N,0}\right)$. Interestingly, the resulting \textit{coupling coefficients}, $\kappa_m=\frac{\kappa}{2}\sqrt{m(N+1-m)}$, are identical to the coupling coefficients required to implement a discrete fractional Fourier transformer \cite{Weimann}. 
In what follows we omit the term $N \beta \delta_{n,m}$ since it only contributes to the dynamics as a global phase. \\
Using the matrix elements $\hat{H}_{n,m}$ one can readily see that for $N$ photons, the $(N+1)$ possible states obey the set of coupled equations
\begin{align}
\mathrm{i} \frac{d}{dz} |n) &= \kappa_{n} |n-1)+\kappa_{n+1} |n+1),
\end{align}
with $n=0,...,N$. In Fig.~(\ref{fig:coupler} b) we show a diagrammatic representation of the ($N+1$) states arising when the state $|m)$ is fed into the BS shown in Fig.~(\ref{fig:coupler} a).
In fact, by starting with the initial state $\ket{\psi(0)}=|m)$, one can show that the probability amplitude of finding the system in state $|n)$, after a propagation distance $z$, is analytically described by \cite{Weimann}
\begin{align}\label{eq:Green}
\begin{split}
G_{n,m}(z) = &\mathrm{i}^{n-m} \sqrt{\frac{n!(N-n)!}{m!(N-m)!}} \left[ \sin\left( \frac{\kappa z}{2}\right)\right]^{m-n} \\  \times &\left[ \cos \left(\frac{\kappa z}{2} \right) \right]^{N-n-m}  P^{(m-n,N-m-n)}_n(\cos(\kappa z)),
\end{split}
\end{align}
where $P^{(a,b)}_n(x)$ are the Jacobi-Polynomials of order $n$.
To visually illustrate this result, in Fig.~(\ref{fig:Dyn}) we present the $N$-photon dynamics for a BS having $\kappa=1$ excited by the $N=9$ state $\ket{\psi(0)}=|3)$. Fig.~(\ref{fig:Dyn} a) depicts the coupling distribution computed from the matrix elements $\hat{H}_{n,m}$. From this figure it is clear that the coupling distribution among the states exhibits a parabolic shape. This implies that the central states, in this case $\ket{4,5}$ and $\ket{5,4}$, are coupled more strongly than those lying close to the edge of the state diagram, i.e., $\ket{0,9}$ and $\ket{1,8}$. \\
A notable feature of discrete fractional Fourier systems is their capability to perform mirror inversions of the initial states at the propagation distance $z=\pi/\kappa$. Consequently, any initial state $|m)$ will be transformed into its mirror image $|N-m)$. This property follows from the fact that at the particular distance $z=\pi/\kappa$ all probability amplitudes, $G_{n,m}\left(\pi/\kappa\right)$, vanish except for $n=N-m$. Accordingly, for the example shown in Fig.~(\ref{fig:Dyn}~b) $(\kappa=1)$, we see that at $z=\pi$ the photons indeed swap position, $\ket{\psi(0)}=|3)=\ket{3,6}\rightarrow\ket{\psi(\pi)}=|6)=\ket{6,3}$. This intrinsic mirror inversion property of BSs is rather appealing for quantum computing as it can be exploited to perform swap operations over multiphoton states.\\ 
To demonstrate the potential of the present approach in a more practically relevant scenario we consider the excitation of the waveguide BS by a two-mode squeezed vacuum state (TMSVS) given by $\ket{\xi}=\sqrt{1-|\xi|^{2}}\sum_{n=0}^{\infty}\xi^{n}\ket{n,n}$. Here, $\xi$ denotes the squeeze parameter and it is restricted to the unit circle in the complex plane, $0\le |\xi|\le 1$ \cite{Gerry}. Quite interestingly, in this case all the states $\sqrt{1-|\xi|^{2}}\xi^{n}\ket{n,n}$ evolve paralelly and independently from each other undergoing quantum walks in the photon number space, see Fig.~(\ref{fig:parallel}).\\ 
In the optical domain, continuous time quantum random walks  are the quantum analog of classical random walks in which initially localized photons coherently evolve into multiple spatial-mode superpositions of discrete systems, e.g. integrated waveguide lattices or cascaded bulk BSs \cite{Broome,Perets,Svozilik,Grafe,Grafe2}. Naturally, in such configurations the maximum number of steps that the walker can perform is determined by the number of sites contained in the lattices. 
In our protocol, distinct QWs are produced simultaneously and the corresponding walkers perform a different number of steps which depends on the number of photons involved in each process. Moreover, the probability of occurrence for each QW is weighted by the initial multiphoton probability amplitudes, $\sqrt{1-|\xi|^{2}}\xi^{n}$. We point out that the observation of these types of processes is nowadays possible utilizing bright parametric-down-conversion sources \cite{Gerry} in combination with photon-number resolving detectors \cite{Lita}. In fact, recently two-mode quantum light with a mean photon number of 50 and a maximum number of 80 photons for each of the two modes has been reported in \cite{Harder}.\\
In addition to the mirror inversion property, discrete fractional Fourier transformers present the very unique property that at $z=\pi/2\kappa$, the distance at which the usual DFT occurs, any \textit{single-point} excitation $|m)$ evolves into the $m$-th eigenstate, $u^{(m)}_n$, of the Hamiltonian $\hat{H}_{n,m}$ up to some local phases, $G_{n,m}\left(z=\frac{\pi}{2\kappa}\right)= \mathrm{i}^{n-m} u^{(m)}_n$ \cite{Weimann}, where $u^{(m)}_n=2^{-\frac{N}{2}+n}\sqrt{\frac{n!(N-n)!}{m!(N-m)!}}P^{(m-n,N-m-n)}_n(0)$, and the Jacobi polynomials are evaluated at the origin. 
Based on this property, we formulate an $N$-photon suppression law \cite{Tichy1,Tichy2,Tichy3} for 50/50 waveguide BSs acting as DFrFT in the $N$-photon subspace.
Interestingly, the first manifestation of such a photon suppression law occurring in BSs, that is, suppression of two photon states in a BS, was observed by Hong, Ou, and Mandel \cite{Hong}. Their experiment consisted of launching two indistinguishable photons onto opposite sides of a 50/50 BS and they observed that the two photons always emerged from the same output port with the same probability, while the state $|1)=\ket{1,1}$ was completely absent or suppressed. 
In a similar fashion, here, we provide an answer to the question: given the $N$-photon input state $|m)=\ket{m,N-m}$ for a 50/50 waveguide BS, what output-states $|n)=\ket{n,N-n}$ will exhibit zero probability amplitude?
\begin{equation}\label{eq:zeros}
G_{n,m}\left(z=\frac{\pi}{2\kappa}\right)=\mathrm{i}^{n-m} u^{(m)}_n \stackrel{!}{=}0. 
\end{equation}
A close inspection of the Jacobi polynomials contained in $u^{(m)}_n$ reveals that the only possibility for Eq.~(\ref{eq:zeros}) to be zero is when the following expression becomes zero
\begin{equation} \label{eq:supp-law}
Q^{(N)}_{n,m}:=\sum_{k=0}^n (-1)^k \binom{m}{k}\binom{N-m}{n-k}=0.
\end{equation}
Quite interestingly, in reference \cite{Gy} some particular solutions of Eq.~(\ref{eq:supp-law}) have been reported, although a general solution appears to be elusive. First, we denote the solutions of \eqref{eq:supp-law} as the 3-tuple $F=(N,n,m)$, where $N$ is the total number of photons, $n$ indicates the suppressed output state $|n)$, and $m$ the input state $|m)$. 
For the case $N =$ even and the initial state $|m)=|N/2)$, one can show that
\begin{equation} \label{eq:trivial}
Q^{(N)}_{n,N/2}=0 \Leftrightarrow n \mbox{ is odd.}
\end{equation}
Therefore, we can choose the 3-tuple $F_0=\left(2k,2k'+1,k \right)$, for any integers $k\geq 1,k'\geq0$, as the first family of solutions to \eqref{eq:supp-law}. Accordingly, by sending an equal number of photons into both waveguides will render the odd final states to be suppressed. 
For the special case of $N=2$ and the initial state $|1)=\ket{1,1}$, that is, the Hong-Ou-Mandel case, \eqref{eq:trivial} predicts that $Q^{(2)}_{1,1}=0$, or equivalently $G_{1,1}\left(z=\frac{\pi}{2\kappa}\right)=0$, which indicates that the state $|1)=\ket{1,1}$ is completely suppressed.\\ 
Before considering the case $N=$ odd, it is important to note that the eigenstates $u^{(m)}_n$ satisfy the symmetry relations $u^{(m)}_n = (-1)^{n-m}u^{(n)}_m = (-1)^{n} u^{(N-m)}_n$.
From these relations it is clear that, when $(N,n,m)$ satisfies \eqref{eq:supp-law}, the combinations $(N,N-n,m)$, $(N,n,N-m)$, $(N,m,n)$, $(N,N-m,n)$, $(N,m,N-n)$, $(N,N-n,N-m)$ and $(N,N-m,N-n)$ also satisfy \eqref{eq:supp-law}. It is worth emphasizing that these solutions are implicitly contained in $F_0$. \par
For the case $N=$ odd there is no general rule to predict state suppressions. There are, however, certain \textit{anomalous} (i.e. not contained in $F_0$) photon numbers $N_a$, both even and odd, where 3-tuples $(N_a,n,m)$ with $n,m\neq N_a/2$ satisfy \eqref{eq:supp-law}. The ordered list of these \textit{anomalous} photon numbers up to 67 is $N_a = 9,16,17,22,25,33,34,36,41,49,57,64,65,66,67$ \cite{Gy}.
Notice, the odd numbers contained in this list can be obtained by taking $N_{a}=8k+1$ with $k\geq 1$. Moreover, one can show that by taking $m=4k-1$ renders $Q^{(8k+1)}_{2k,4k-1}=0$. This implies that a total number of photons $N_{a}=8k+1$ prepared in the state $|4k-1)=\ket{4k-1,4k+2}$ will produce an output in which the states $|2k)=\ket{2k,6k+1}$ will be absent.\\
As an example we consider the case $k=1$, which corresponds to the smallest \textit{anomalous} photon number $N_a=9$. Accordingly, for $N_a=9$ we have $Q^{(8k+1)}_{2k,4k-1}=Q^{(9)}_{2,3}=0$, which implies the combination $(N_{a},n,m)=(9,2,3)$. Hence, by launching the initial state $|3)=\ket{3,6}$ will render the output state $|2)=\ket{2,7}$ to be suppressed. These effects are shown in Fig.~(\ref{fig:Dyn}~b), where it is clear that at $z=\pi/2$ (along the left dashed white line) the states $|2)=\ket{2,7}$ and $|7)=\ket{7,2}$ are suppressed. Furthermore, for $N_{a}=9$ we can use the aforementioned symmetry properties to obtain the combinations $(n,m)=(7,3),(2,6),(3,2),(6,2),(3,7),(7,6)$, and $(6,7)$. In detail, the pair $(n,m)=(6,7)$ indicates that by launching the 9-photon state $|7)=\ket{7,2}$ into a  50/50 waveguide BS, the state $|6)=\ket{6,3}$ will be suppressed at the output.\par
In addition to the family of solutions $F_{0}$ there exist four more families given by \cite{Gy}
\begin{align}\label{eq:11}
F_1&=\left(8k+1,2k,4k-1 \right),\mbox{ } k\geq 1\\\label{eq:12}
F_2&=\left((k+2)^2,\frac{(k+1)(k+2)}{2},2\right), \mbox{ } k\geq 1\\\label{eq:13}
F_3&=\left(3k^2+8k+6,\frac{(k+1)(3k+2)}{2},3 \right),\mbox{ } k\geq 1\\\label{eq:14}
F_4&=\left(3k^2+10k+9,\frac{(k+1)(3k+4)}{2},3 \right),\mbox{ } k\geq 0.
\end{align}
Using these solutions one can compute most combinations $(n,m)$ of suppressed output states $|n)$ when the system is excited by state $|m)$, for $N\leq 100$. In this range, however, the anomalous photon numbers $N_a = 36, 66, 67, 98$ display additional suppressions, that are not predicted by any of the known families of solutions $F_{0,1,2,3,4}$. 
For example,  by taking $k=4$ in $F_2$ yields the set $(36,15,2)$, however, there exists an additional set of solutions $(36,14,5)$, which do not follow from applying the symmetry relations to $(36,15,2)$.\\
\section{SUMMARY}
In this paper, based on the DFrFT properties of 50/50 waveguide BSs, we have derived a suppression law, \eqref{eq:supp-law}, for $N$-photon states.
This result can be understood as the N-photon generalization of the Hong-Ou-Mandel effect \cite{Carolan,JPhysB}. Interestingly, due to the fact that the derived suppression law only holds for indistinguishable photons, it constitutes an experimentally feasible test for certifying true multiphoton interference, such as in the boson-sampling problem \cite{lund2017}. Extending this multiphoton lattice-like formalism to systems comprising many modes \cite {Rechtsman} may open the door to new applications. We stress that the results presented in this work are not restricted to using integrated waveguide BSs and one can use bulk BSs with different splitting ratios. Finally, the family of solutions presented here --together with the rapid development of highly efficient photon sources and detectors-- constitute a fundamental step towards the efficient generation of multiphoton states via many-particle interference.
\vspace{-1cm}
\section*{Acknowledgements}
RJLM thankfully acknowledges financial support by CONACYT-M\'exico under the project CB-2016-01/284372, and by DGAPA-UNAM under the project UNAM-PAPIIT IA100718.

\end{document}